\documentclass[10pt,a4paper,twocolumn,english,aps,prl,twocolumn,showpacs,floatfix,superscriptaddress,10pt]{revtex4-1}
\usepackage[T1]{fontenc}
\usepackage[latin9]{inputenc}
\usepackage{color}
\usepackage{babel}
\usepackage{amstext}
\usepackage{amssymb}
\usepackage{graphicx}
\usepackage[unicode=true,
 bookmarks=true,bookmarksnumbered=false,bookmarksopen=false,
 breaklinks=false,pdfborder={0 0 0},backref=false,colorlinks=true]
 {hyperref}
\hypersetup{pdftitle={Experimental evidences of a large extrinsic spin Hall effect in AuW alloy.},
 pdfauthor={Piotr Laczkowski}}
\usepackage{breakurl}

\makeatletter

\@ifundefined{textcolor}{}
{%
 \definecolor{BLACK}{gray}{0}
 \definecolor{WHITE}{gray}{1}
 \definecolor{RED}{rgb}{1,0,0}
 \definecolor{GREEN}{rgb}{0,1,0}
 \definecolor{BLUE}{rgb}{0,0,1}
 \definecolor{CYAN}{cmyk}{1,0,0,0}
 \definecolor{MAGENTA}{cmyk}{0,1,0,0}
 \definecolor{YELLOW}{cmyk}{0,0,1,0}
}

\@ifundefined{date}{}{\date{}}
\makeatother

\begin{document}

\title{Experimental evidences of a large extrinsic spin Hall effect in AuW
alloy.}

\author{P. Laczkowski}

\affiliation{Unité Mixte de Physique CNRS/Thales and Université Paris-Sud 11,
91767 Palaiseau, France }

\author{J.-C. Rojas-Sánchez }

\affiliation{Unité Mixte de Physique CNRS/Thales and Université Paris-Sud 11,
91767 Palaiseau, France }

\author{W. Savero-Torres }

\affiliation{INAC/SP2M, CEA-Université Joseph Fourier, F-38054 Grenoble, France}

\author{H. Jaffrès }

\affiliation{Unité Mixte de Physique CNRS/Thales and Université Paris-Sud 11,
91767 Palaiseau, France }

\author{N. Reyren }

\affiliation{Unité Mixte de Physique CNRS/Thales and Université Paris-Sud 11,
91767 Palaiseau, France }

\author{C. Deranlot }

\affiliation{Unité Mixte de Physique CNRS/Thales and Université Paris-Sud 11,
91767 Palaiseau, France }

\author{L. Notin }

\affiliation{INAC/SP2M, CEA-Université Joseph Fourier, F-38054 Grenoble, France}

\author{C. Beigné }

\affiliation{INAC/SP2M, CEA-Université Joseph Fourier, F-38054 Grenoble, France}

\author{A. Marty }

\affiliation{INAC/SP2M, CEA-Université Joseph Fourier, F-38054 Grenoble, France}

\author{J.-P. Attané }

\affiliation{INAC/SP2M, CEA-Université Joseph Fourier, F-38054 Grenoble, France}

\author{L. Vila}

\affiliation{INAC/SP2M, CEA-Université Joseph Fourier, F-38054 Grenoble, France}

\author{J.-M. George }

\affiliation{Unité Mixte de Physique CNRS/Thales and Université Paris-Sud 11,
91767 Palaiseau, France }

\author{A. Fert}

\affiliation{Unité Mixte de Physique CNRS/Thales and Université Paris-Sud 11,
91767 Palaiseau, France }

\date{\today}
\begin{abstract}
We report an experimental study of a gold-tungsten alloy (7\% at.
W concentration in Au host) displaying remarkable properties for spintronics
applications using both magneto-transport in lateral spin valve devices
and spin-pumping with inverse spin Hall effect experiments. A very
large spin Hall angle of about 10\% is consistently found using both
techniques with the reliable spin diffusion length of $2\, nm$ estimated
by the spin sink experiments in the lateral spin valves. With its
chemical stability, high resistivity and small induced damping, this
AuW alloy may find applications in the nearest future.
\end{abstract}
\maketitle

The spin Hall effect (SHE) \citep{Dyakonov1971459,Hirsch1999} is
an emerging route for spintronics since it allows for conversion of
charge into pure spin currents or \textit{vice-versa} through the
direct or inverse SHE \citep{Inoue2005} in non-magnetic materials.
For instance, pure spin currents without a net charge current can
be produced this way but require materials with strong spin-orbit
(SO) interaction, beyond solely Pt \citep{saitoh:182509,kimura2007roo,Vila2008PRL,Ralph_PtReview_2012}
or Pd \citep{Ando2010a}. The conversion ratio between the charge
and spin currents is called the spin  Hall angle ($\theta_{SHE}$).
The SHE was first observed in semiconductor materials using optical
methods \citep{kato2004observation,wunderlich2005experimental}. More
recently, reported values of a few percent in metals revived the subject
\citep{saitoh:182509} and led to an investigation of several metals
and alloys with reported spin Hall angle value as large as 30\% \citep{Liu_Science_2012_ST_SHE_Ta}.
Despite the large dispersion of spin Hall angle and a long standing
debate about the results \citep{Ralph_PtReview_2012,Niimi2013,Miron2011},
experiments showed that the spin current produced by the spin-orbit
effects can be successfully used in order to electrically control
magnetization \textit{via} spin-torque switching of ferro-magnets
\citep{Miron2011,Ralph_PtReview_2012} as well as precession in spin-torque
ferromagnetic resonance \citep{Liu_STFMR_2011}. Additional proposition
are the stirring effect \citep{Pershin_APL_2009}, the spin Hall effect
transistors along  gated semiconductor channels \citep{wunderlich2010spin},
or the charge production for spincaloritronics \citep{bauer_Nature_2012}.
An important challenge for applications is to find materials with
an efficient spin to charge current conversion. Large $\theta_{SHE}$
of intrinsic origin were predicted theoretically \citep{Tanaka2008,Kontani2009}
and confirmed experimentally \citep{Morota_PRB_2011} in 4d and 5d
transition metals. The extrinsic SHE however, through either the skew-scattering
\citep{Smit1958} or the side jump \citep{Berger1970} mechanism,
allows even better control of the $\theta_{SHE}$ by tuning the impurity
concentration in a host material \citep{Fert1981} and taking advantage
of resonant scattering on impurity levels split by the spin orbit
interaction \citep{FertLevy_PRL_2011}. Therefore, even larger effects
in doped materials with suitable host and impurities with large SO
interaction could be foreseen, as was recently found in CuIr \citep{Niimi2011},
CuBi \citep{Niimi_PRL_CuBi_2012} or CuPb \citep{Niimi_PRB_2014}.
These materials can act as building blocks for new spintronics applications
such as spin current injectors or detectors as demonstrated through
CuIr Magnetic Tunnel Junctions (MTJ) \citep{Yamanouchi2013}, write
heads \citep{Datta2012} or Giant-SHE-MTJs \citep{Manipatruni2013}.

In this letter we present analyses of a metallic AuW alloy (7\% at.
W concentration in Au host - controlled using particle-induced X-ray
emission technique) exhibiting a large $\theta_{SHE}$, using two
complementary methods: lateral spin valves and inverse spin Hall effect
(ISHE) along with spin pumping. The first method consists of the lateral
spin valves (LSV) experiments allowing characterization of materials
with small spin diffusion length \citep{Kimura2005PRB}. 

\begin{figure}
\begin{centering}
\includegraphics[width=8.5cm]{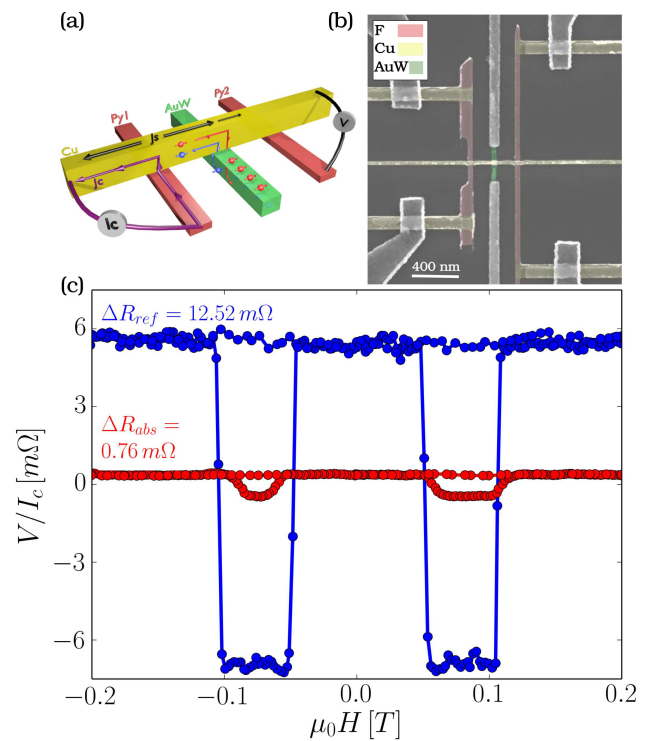}
\par\end{centering}

\caption{\textit{\label{Fig1_SpinSink}(a) Schematic representation and (b)
a Scanning Electron Microscope image of a lateral spin valve fabricated
using multi-angle shadow evaporation technique with }$AuW_{7\%}$\textit{
nano-wire inserted in-between two Py ferromagnets. Spin signal recorded
at $T=11\, K$ for nano-structures with (red) and without (blue) }$AuW_{7\%}$\textit{
nano-wire insertion.}}
\end{figure}

Figure \ref{Fig1_SpinSink} represents (a) a sketch and (b) a scanning
electron microscope image of a typical device fabricated using the
multi-angle nano-fabrication method \citep{Laczkowski2011APEX}, where
the colors red, yellow, and green represents ferromagnetic, non-magnetic
and SHE material respectively. First, the middle $AuW_{7\%}$ wire
is deposited on the $SiO_{2}$ substrate using Physical Vapor Deposition
(PVD) and a lift-off technique, followed by the nano-fabrication of
a $Py/Cu$ lateral spin-valve. In between these two steps the middle
wire surface is cleaned using Ar ion-milling. The Py, Cu and $AuW_{7\%}$
nano-wires are $20\, nm$, $70\, nm$ and $20\, nm$ thick respectively.
Their width is fixed to $50\, nm$ with an exception of the $100\, nm$
wide $AuW_{7\%}$. The distance separating the ferro-magnets in the
presented device is $L=400\; nm$ (from center to center of ferro-magnets).

Non-local measurements have been performed, using a standard Lock-in
amplifier technique at $f=79\, Hz$, for two types of nano-structures.
The first one as a reference is a regular non local device \citep{Laczkowski2011APEX}
including a Cu channel and two ferromagnetic Py electrodes whereas
the second one includes a $AuW_{7\%}$ wire inserted in between the
ferromagnetic electrodes {[}Fig. \ref{Fig1_SpinSink}(a-b){]}. In
these experiments a spin accumulation is created in the LSV by passing
a partly spin-polarized charge current through a ferromagnetic/non-magnetic
interface ($Py1/Cu$) {[}Fig. \ref{Fig1_SpinSink}(a){]}. The spin
accumulation diffuses in both directions from this interface in Cu
giving rise to spin currents. On the right-hand side of the LSV there
are only pure spin currents without a net charge flow ($J_{s}$ black
arrows ) which are partially absorbed by the $AuW_{7\%}$ nano-wire.
The voltage is measured across the second $Py2/Cu$ interface in order
to probe spin currents arriving at this interface. The amplitude of
the spin signal for reference and absorption devices was measured
to be $\Delta R_{ref}=12.52\, m\Omega$ and $\Delta R_{abs}=0.75\, m\Omega$
respectively which yields an absorption of $94\%$ (independent of
temperature) {[}Fig. \ref{Fig1_SpinSink}(c){]}. 

These experiments allow the experimental evaluation of the spin diffusion
length of of AuW7\% to be determined as long as the absorption ratio
is known. The resistivity of the inserted material, $\rho_{SHE}$,
can be directly measured, however the spin diffusion length needs
to be calculated. For the general case of the LSV, using a 1D model
\citep{Kimura2005PRB,Jaffres_PRB_2010}, the ratio of the spin signal
amplitude with the middle wire to the spin signal amplitude without
this nano-wire equals %
\footnote{Presented formula is a simplified version of the equation from \citep{Kimura2005PRB}.%
}:
\begin{equation}
\begin{array}{c}
\frac{\Delta R_{abs}}{\Delta R_{ref}}=\frac{(R_{N}+\exp(L/l_{sf}^{N})(R_{F}+R_{N}))R_{AuW}}{R_{N}(-R_{N}+R_{AuW})+\exp(L/l_{sf}^{N})(R_{F}+R_{N})(R_{N}+R_{AuW})}\end{array}\label{eq:Rshe_calculatiion}
\end{equation}

Here $R_{N(F)}=\rho_{N(F)}l_{sf}^{N(F)}/A_{N(F)}(1-P_{F}^{2})$ stands
for spin resistances, where: $A_{N}=w_{N}\times t_{N}$, $A_{F}=w_{F}\times w_{N}$,
$\rho_{N(F)}$, $l_{sf}^{N(F)}$, $t_{N}$, $w_{N(F)}$, $A_{N(F)}$
are the resistivity, the spin diffusion length, the thickness, the
width and the effective cross sectional area respectively. In the
presented notation subscripts F and N correspond to the ferromagnetic
and non-magnetic material respectively. The spin resistance of the
inserted $AuW_{7\%}$ nano-wire is: $R_{AuW}=\frac{\rho_{AuW}\, l_{sf}^{AuW}}{w_{N}\, w_{AuW}tanh(t_{AuW}/l_{sf}^{AuW})}$ 

This allows for $l_{sf}^{AuW}$ of the AuW nano-wire to be extracted
knowing material characteristic parameters of the LSV: AuW resistivity
$\rho_{AuW_{7\%}}=570\,\Omega.nm$ and thickness $t_{AuW_{7\%}}=20\, nm$.
Remaining LSVs parameters are: $\rho_{Cu}^{11K}=55\,\Omega.nm$, $\rho_{Py}^{11K}=118\,\Omega.nm$,
and the characteristic spin transport parameters: $l_{sf}^{Cu}=320\, nm$
and $P_{eff}=0.36$, which were taken from our previous experiments
\citep{laczkowski2012_PhD,rojas2013plane,motzko2013pure}. These results
are in agreement with what can be found in literature for similar
structures \citep{Bass2007}. This analysis leads to the spin diffusion
length of $AuW_{7\%}$ to be $l_{sf}^{AuW7\%}=1.9\, nm$.

Moreover, one can also perform SHE experiments in the very same structure.
A spin current is injected electrically by using $Py1/Cu$ interface
as presented above. The spin current absorbed by the $AuW_{7\%}$
nano-wire is then converted into a transverse charge current \textit{via}
the SHE and thus leads to an electric voltage signal. The external
magnetic field is swept along the non-magnetic Cu channel while a
voltage drop is measured on the $AuW_{7\%}$ nano-wire edges. 

\begin{figure}
\begin{centering}
\includegraphics[width=8.5cm]{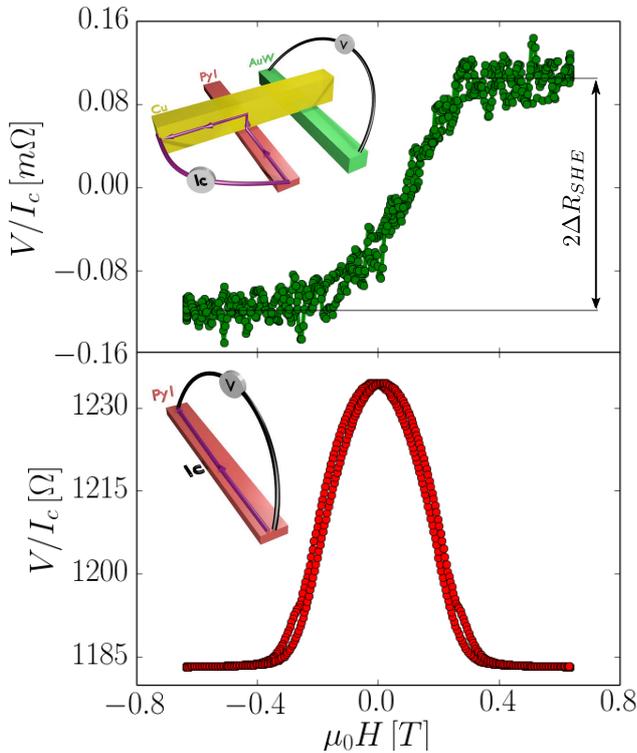}
\par\end{centering}

\caption{\textit{\label{fig:SHE_LSVs}(a) Inverse-SHE measurements in $AuW_{7\%}$
$100\, nm$ width nano-wire with (b) Anisotropic-Magnetoresistance
signal of the ferromagnetic injector, both measured at $T=10\, K$.
External magnetic field was applied along the non-magnetic channel
direction. The saturation field in both figures correspond. (Insets)
Schematic representation of used measurement configurations.}}
\end{figure}

Figure \ref{fig:SHE_LSVs} shows experimental data-points obtained
at $T=11\, K$ for the Inverse-SHE when the voltage is measured at
the edge of the AuW wire and Anisotropic Magneto-resistance (AMR)
in 2 contacts at the edge of the Py injector. Probe configurations
are represented schematically in corresponding insets. A clear non-local
signal is observed yielding $\Delta R_{SHE}=0.1\, m\Omega$. Its amplitude
is a roughly linear function of the magnetic field in a region below
the saturation field $\mu_{0}H_{sat}=0.38\, T$ which corresponds
to the situation where the magnetization of the ferromagnetic injector
(Py1) is fully parallel to the non-magnetic Cu channel.

\begin{figure}
\begin{centering}
\includegraphics[width=7cm]{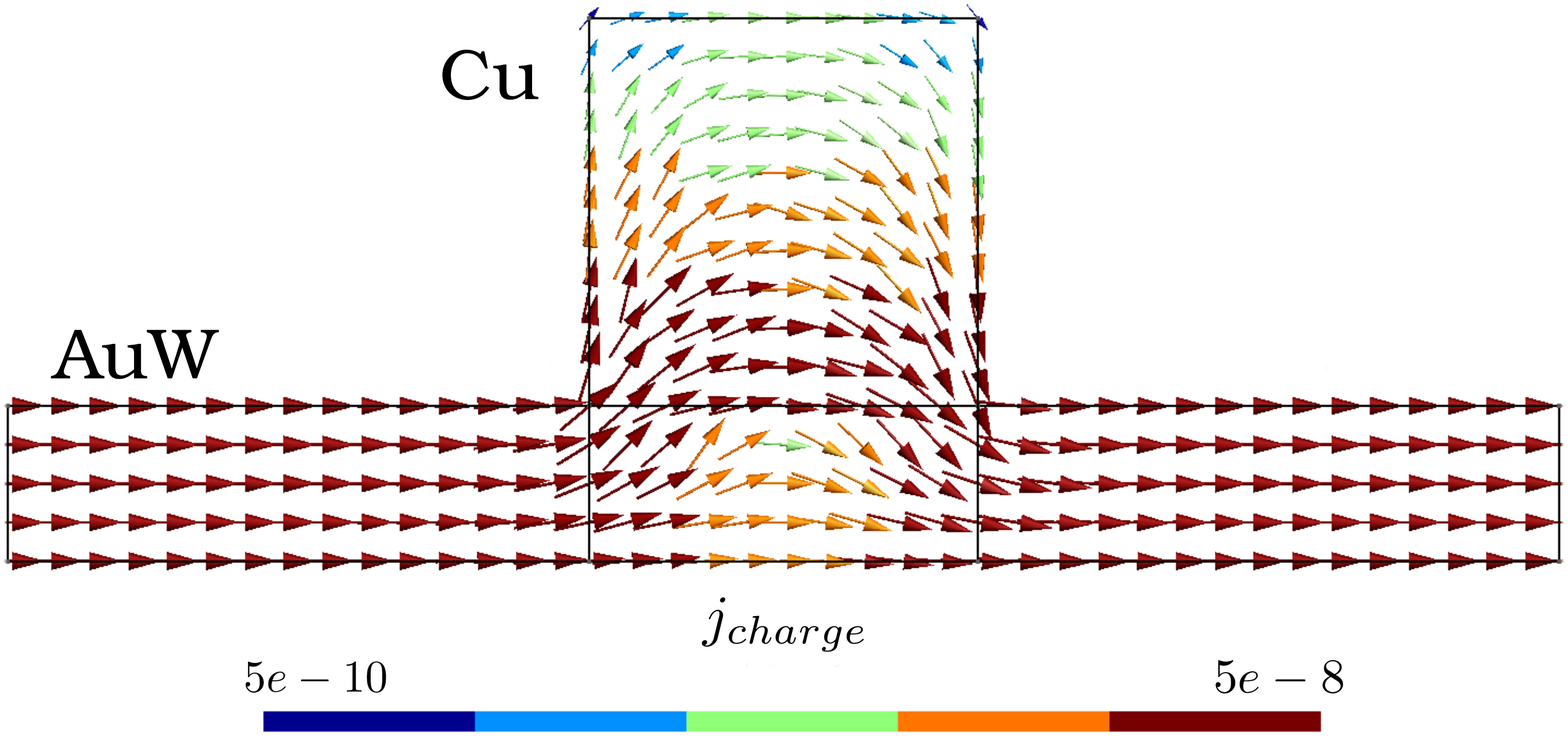}
\par\end{centering}

\caption{\textit{\label{fig:Shunting effect}Cross section of the $AuW/Cu$
interface charge current distribution using FEM simulations represented
in the logarithmic scale.}}
\end{figure}

It is known that evaluating the spin Hall angle with a simple 1D model
leads to underestimation of the $\theta{}_{SHE}$ due to the shunting
effect \citep{Niimi2011,Niimi_PRL_CuBi_2012}. We then performed a
more sophisticated analysis using a Finite Element Method (FEM) simulation
\citep{laczkowski2012_PhD}. The result of this analysis shows that
the spin Hall angle $\theta_{SHE}^{AuW7\%\,(3D)}=10\%$ is larger
than the one obtained with the 1D model ($\theta_{SHE}^{AuW7\%\,(1D)}=1\%$)
\citep{Takahashi_SHE_equation}. Two effects contribute to underevaluation
of the $\theta_{SHE}^{AuW7\%}$ with the 1D model. The main effect
is due to shunting by the Cu layer of a part of the ISHE current produced
by the spin current absorption in AuW nano-wire. The current distribution
is quite inhomogeneous and therefore one should take into account
only a shunt by a part of the Cu layer in the same order of magnitude
than about half the contact width. This leads to the 1D model correction
factor of about 12 (related directly with the $AuW/Cu$ resistivity
ratio). This can be view in figure \ref{fig:Shunting effect} where
90\% of the current extends on about $18\, nm$ in Cu layer. This
emphasizes the 1D model limitation with these aspect ratios and the
requirement to use a finite element model in order to analyze the
data correctly. The second effect is linked to the spreading of the
spin accumulation over the sides of the contacts in the SHE material
mainly when the spin diffusion length of the SHE materials is long,
compared to the width . In our case with a spin diffusion length
of about $2\, nm$ compared to the thickness of $20\, nm$, this effect
should be negligible giving a posteriori justification of the derivation
of the $l_{sf}^{AuW_{7\%}}$ with 1D model {[}Eq. \ref{eq:Rshe_calculatiion}{]}. 

We turn now to the second experiment performed at room temperature:
spin pumping and ISHE where the ferromagnetic resonance (FMR) is exploited.
The resistivity variation from 10K up to 300K for AuW7\% is $\Delta\rho/\rho\sim3\%$,
almost insensitive to temperature due to the dominant role of the
scattering by impurities. We thus expect $l_{sf}^{AuW_{7\%}}$ (as
well as $\theta_{SHE}^{AuW_{7\%}}$) to be independent of the temperature
and that the two measurement results can be directly compared. We
have grown by sputtering a $\|Py(15\, nm)/AuW_{7\%}(30\, nm)$ bilayer,
and a $\|Py(15)/Al(7\, nm)$ reference sample. In a usual FMR experiment,
a microwave is applied at a given frequency and power with a magnetic
field strength $h_{rf}$. Additionally, a perpendicular $dc$ magnetic
field $\mu_{0}H_{dc}$ is swept. At the resonance condition, when
$H_{dc}$ is parallel to the film plane, a pure $dc$ spin current
is be injected from the Py layer to the $AuW_{7\%}$ layer. This is
the spin pumping effect \citep{Tserkovnyak2005} which can be detected
by the increase of the damping constant in the bilayer with respect
to the single Py reference sample. Due to the ISHE in the $AuW_{7\%}$
layer, the injected $dc$ spin current is converted into a $dc$ charge
current, which, in turn, can be detected as a transverse $dc$ voltage. 

\begin{figure}
\begin{centering}
\includegraphics[width=8.5cm]{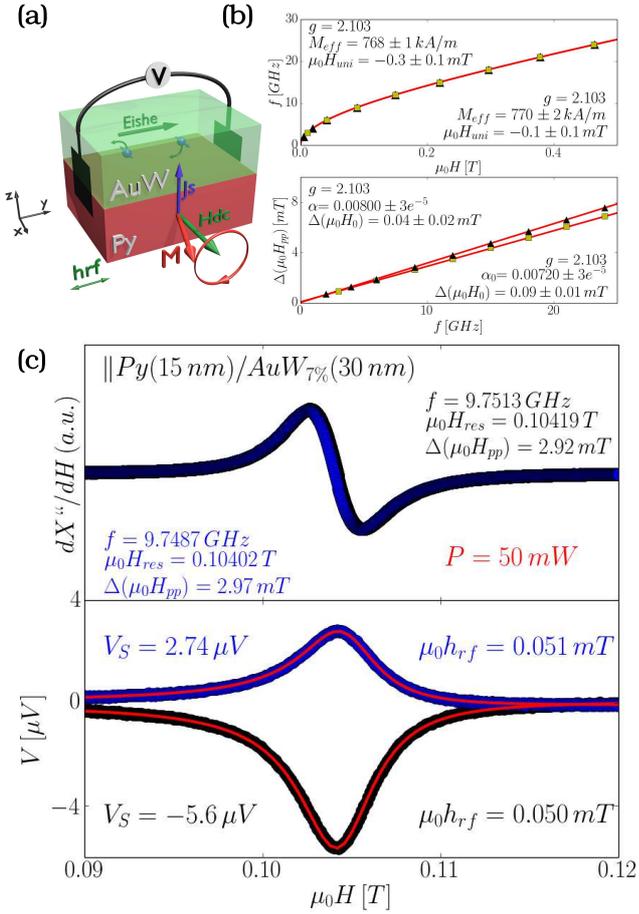}
\par\end{centering}

\caption{\label{fig3:SP-ISHE} \emph{(a) A sketch of the Py($15\, nm$)/AuW$_{7\%}$($30\, nm$)
sample at the resonance condition. A spin current is injected from
the Py layer into the AuW$_{7\%}$ layer. Due to the inverse spin
Hall effect the spin current is converted into a charge current and
detected by measuring the transverse $dc$ voltage. Experimental data
for (b) the frequency dependence of the resonance field (top) and
of the peak-to-peak linewidth (bottom) after FMR spectrum measurements
in a broadband frequency range. Lines represent fits of Kittel's relationship
(top) and a Gilber-type damping contribution (bottom). Fitted values
are displayed in each panel. (c) The FMR spectrum and the $dc$ voltage
measured simultaneously at $9.75\, Ghz$ in a cylindrical resonant
cavity. The $h_{rf}$ value was measured with samples inside the cavity.
Black dots represent the peak-down voltage data in the parallel case,
while blue ones represent the case of sample turned 180 degrees in-plane
(anti-parallel). Lines show for a Lorentzian fits of the voltages.
The symmetrical voltage amplitude $V_{S}$ is displayed for each case.}}
\end{figure}

$_{}$A typical schematic of the spin pumping-ISHE experiment is shown
in Fig. \ref{fig3:SP-ISHE}(a). As described previously \citep{Jain2012,Rojas-Sanchez2013}
we have used a broadband stripe system to accurately determine the
damping constant {[}Fig. \ref{fig3:SP-ISHE}(b){]} and we have used
an X-band cylindrical cavity to measure the FMR spectrum and the voltage
due to the ISHE {[}Fig. \ref{fig3:SP-ISHE}(c){]}. The frequency \emph{f}
\textit{vs}. magnetic resonant field, \emph{$\mu_{0}H_{res}$, }allows
the effective magnetic saturation and in plane anisotropy to be calculated
as displayed in Fig. \ref{fig3:SP-ISHE}(b) top. We can observe that
both samples, the $\|Py/Al$ reference and the $\|Py(15\, nm)/AuW_{7\%}(30\, nm)$
bilayer have the same $M_{eff}$ value and a negligible in-plane magnetic
anisotropy (\emph{$\mu_{0}H_{uni}$}). The linear frequency dependence
of the peak-to-peak linewidth ($\Delta(\mu_{0}H_{pp})$) allows extraction
of the damping constant $\alpha$ and the \emph{f}-independent contribution
$\Delta(\mu_{0}H_{0})$, due to inhomogeneities. A linear behavior
is found on both samples in the whole experimental range {[}Fig. \ref{fig3:SP-ISHE}(b)
bottom{]}. Furthermore, the contribution due to inhomogeneities is
very small, $\Delta(\mu_{0}H_{0})<0.1\, mT$, and a small but clear
increase of the damping constant for $\|Py(15\, nm)/AuW_{7\%}(30\, nm)$
can be observed (with respect to the reference sample). Thus we can
conclude, that the $AuW_{7\%}$ layer acts as a spin sink layer. This
can be verified in the simultaneously $FMR-V_{dc}$ measurement {[}Fig.
\ref{fig3:SP-ISHE}(c){]}, where we can observe a symmetrical Lorentzian
voltage peak at the resonance field.

In order to quantify the spin Hall angle of this material we use all
our experimental values taking advantage of the spin diffusion length
already determined in the previous section (which is invariant of
temperature). The effective spin mixing conductivity is calculated
as \citep{Tserkovnyak2005,Mosendz2010c}: $g_{\text{eff}}^{\uparrow\downarrow}=\mu_{0}M_{eff}{{t}_{F}}({{\alpha}_{F/N}}-{{\alpha}_{F}})/(g{{\mu}_{B}})=5.9\times{{10}^{18}}\, m^{-2}$,
where F(N) stands for Py(AuW$_{7\%}$) layer, $\mu_{0}M_{eff}=0.97\, T$
is the effective demagnetization field, $g$ is the Landé g-factor,
$\mu_{B}$ is the Bohr constant, \emph{$t_{F}$} is $15\, nm$, and
all other values are shown in Fig. \ref{fig3:SP-ISHE}(b). The effective
spin current density injected at the $Py/AuW_{7\%}$ interface, while
considering a transparent interface (no spin decoherence), is given
by \citep{Ando2010a,Mosendz2010c,rojas_arXive_2013}:

\textit{\footnotesize{
\begin{equation}
\ensuremath{J_{\text{S}}^{\text{eff}}=\left(\frac{2e}{\hbar}\right)\frac{g_{\text{eff}}^{_{\uparrow\downarrow}}\gamma^{2}\hbar(\mu_{0}h_{\text{rf}})^{2}}{8\pi\alpha_{\text{F/N}}^{2}}\left[\frac{\mu_{0}M_{eff}\gamma+\sqrt{\left(\mu_{0}M_{eff}\gamma\right)^{2}+4\omega^{2}}}{\left(\mu_{0}M_{eff}\gamma\right)^{2}+4\omega^{2}}\right]}
\end{equation}
}}where \emph{e} is the electron charge, $\hbar$ is the Dirac constant,
$\gamma=g\mu_{B}/\hbar$ and $\omega=2\pi f$. Considering $f=9.75\, GHz$
and $\mu_{0}h_{rf}=0.1\, mT$, we obtain $J_{\text{S}}^{\text{eff}}=3.4\, MA/m^{2}$.
In order to eliminate undesirable effects \citep{Feng2012,Rojas-Sanchez2013}
the $dc$ transverse voltage amplitudes in between two $dc$ magnetic
field directions have been averaged and weighted by $(\mu_{0}h_{rf})^{2}$.
We obtain a value of $1650\, V/T^{2}$. Independently, we have measured
the sheet resistance of the bilayer, $R_{S}=9.95\,\Omega/sq$, which
gives the resistance of $L\times W=2.4\times0.4\, mm^{2}$ sized sample
to be $59.7\,\Omega$. Then the $dc$ charge current weighted by $(\mu_{0}h_{rf})^{2}$
is $I_{C}/(\mu_{0}h_{rf})^{2}=28\, A/T{}^{2}$. The conversion between
the spin current density $J_{S}^{\mathrm{eff}}$ and the charge current
due to the ISHE is expressed by \citep{Ando2010a,Mosendz2010c,rojas_arXive_2013}:
\begin{equation}
I_{C}=-W\theta_{SHE}^{AuW_{7\%}}l_{sf}^{AuW_{7\%}}J_{S}^{eff}tanh[\frac{t_{AuW_{7\%}}}{2l_{sf}^{AuW_{7\%}}}]R_{SML}
\end{equation}
 where $t_{N}=30\, nm$ is the thickness of the $AuW_{7\%}$ layer
and $R_{SML}\leq1$ accounts for the spin flip scattering at the $Py/AuW_{7\%}$
interface \citep{rojas_arXive_2013}. Considering transparent interface
($R_{SML}=1$) and using the previous values one obtains a lower boundary
for the spin Hall angle of $AuW_{7\%}$, $\theta_{SHE}^{AuW_{7\%}}=+10\%$,
where the ``+'' indicates the same SHE sign as in Pt. Interface
spin resistance and spin memory loss parameters are not yet well known
in $Py/Au$, but the integration of such quantities will necessarily
lead to an increase of the determined spin Hall angle \citep{rojas_arXive_2013}.
At this point, we cannot state whether the extrinsic SHE is due to
skew scattering or side jump. Additional experiments are in progress
to elucidate this question. In this study we have also verified that
$\theta_{SHE}^{Au}$ for pure Au is around $0.4\%$, which is close
to the value reported elsewhere \citep{Mosendz2010c}. Despite of
the small damping enhancement, comparable with pure Au and Ta, we
have estimated a large $\theta_{SHE}^{AuW_{7\%}}$ which makes this
 alloy attractive for applications which require a low damping constant.

In summary we have successfully developed a gold alloy with a heavy
element in order to increase its spin Hall angle. We have shown by
complementary studies, that the spin diffusion length of $AuW_{7\%}$
is $1.9\, nm$ and its spin Hall angle is at least $10$\%. This value
is more than one order of magnitude higher than in a pure gold and
can probably be further increased with higher impurity concentrations.
The large spin Hall angle in $AuW_{7\%}$ makes this material very
attractive for future developments of spintronics devices, which incorporate
non magnetic materials in order to detect or to generate pure spin
currents. In particular, its stability at room temperature and its
chemical inertia are technologically advantageous compared to other
materials. Future research should investigate the spin memory loss
effect at the interface, to determine the dominant scattering mechanism
and optimize the W concentration in order to maximize the spin/charge
current conversion efficiency. 
\begin{acknowledgments}
We acknowledge U. Ebels, S. Gambarelli, and G. Desfonds for technical
support with the FMR measurements. This Work was partly supported
by the French Agence Nationale de la Recherche (ANR) through projects
SPINHALL (2010-2013) and SOSPIN (2013-2016). 
\end{acknowledgments}

\bibliographystyle{aipnum4-1}

\end{document}